\documentclass[a4paper,11pt]{article}
\usepackage{jheppub} 

\title{\boldmath Note on the local calculation of decoherence of quantum superposition in the static black holes}

\author[a,1]{Ran Li\note[1]{Corresponding author.},}

\author[a]{Zhong-Xiao Man,}

\author[b]{Jin Wang}

\affiliation[a]{Department of Physics, Qufu Normal University, Qufu, Shandong 273165, China}

\affiliation[b]{Department of Chemistry, Physics and Astronomy, Stony Brook University, Stony Brook, NY 11794, USA,}

\emailAdd{liran@qfnu.edu.cn}
\emailAdd{zxman@qfnu.edu.cn}
\emailAdd{jin.wang.1@stonybrook.edu}

\abstract{
We investigate the decoherence of a quantum spatial superposition of a static particle in Schwarzschild and Reissner-Nordstr\"{o}m black holes. By treating the particle as a localized classical source coupled to a quantum scalar field, we reformulate the decoherence process in the Danielson-Satishchandran-Wald (DSW) gedankenexperiment through coherent state generation and derive the local expression for the decoherence functional in terms of the Wightman function. In the long-time limit, the decoherence rate is shown to be characterized by the low-frequency behavior of the Wightman function. We then employ the asymptotic matching method to calculate the analytical expressions of the Wightman functions in the Boulware, Unruh, and Hartle-Hawking vacua. We show that the decoherence behavior depends on the quantum state of the environmental field. While the Boulware vacuum gives vanishing decoherence for a static superposition, the thermal effects associated with Hawking radiation in the Unruh and Hartle-Hawking vacua can induce nonvanishing decoherence. }

\begin{document}

\maketitle
\flushbottom

\section{Introduction} 
\label{sec:intro}

Quantum superposition and quantum coherence are among the most fundamental features of quantum theory \cite{weinberg2015lectures}. In realistic situations, no physical system can be perfectly isolated from its environment. The unavoidable interaction between a quantum system and environmental degrees of freedom leads to the progressive suppression of quantum coherence, a phenomenon known as decoherence \cite{Hornberger:2008xkz}. Therefore, it has been extensively investigated in the framework of open quantum systems \cite{Breuer2002,Giulini1996}. Decoherence has been regarded as an important mechanism for understanding the collapse of quantum states in measurement processes, as well as the emergence of classical behavior from quantum systems \cite{Zurek:1991vd,Zurek:2003zz,Schlosshauer:2003zy}.

Understanding decoherence in gravitational settings is of particular importance for the quantum nature of gravitation and spacetime. Although a complete theory of quantum gravity remains elusive, important insights may still be obtained by studying quantum phenomena in the gravitational and related backgrounds \cite{Bassi:2017szd,Anastopoulos:2021jdz}. In this respect, decoherence induced by gravitational and relativistic effects has attracted increasing attention, including gravity-induced decoherence \cite{Blencowe:2012mp,Pikovski:2013qwa}, acceleration-induced decoherence \cite{Nesterov:2020exl,Xu:2023tdt,Li:2026pxg}.

Recently, a decoherence mechanism proposed by Danielson, Satishchandran, and Wald (DSW) suggests that a quantum spatial superposition of a static particle nearby the black hole horizon can decohere due to environmental field fluctuations associated with horizon physics and Hawking radiation \cite{Danielson:2022tdw,Danielson:2022sga,Danielson:2024yru}. The presence of black hole Killing horizon can inevitably destroy the quantum 
coherence by acquiring ``which-way" information about the superposition through the soft particle (photons or gravitons) radiation crossing the horizon. The DSW decoherence effect has attracted significant attention in recent studies \cite{Gralla:2023oya,Wilson-Gerow:2024ljx,Biggs:2024dgp,Li:2024guo,Li:2024lfv,Li:2025vcm,Kawamoto:2025kfu,Fahn:2025fxl,Fahn:2025jxh,Batista:2026otr,Biggs:2026zlp}. See also Refs.~\cite{Danielson:2025iud,Kudler-Flam:2025yur,Danielson:2025aji,Fahn:2026rei,Gallock-Yoshimura:2026vku} for related quantum information implications.

The original analysis of such decoherence effects \cite{Danielson:2022tdw,Danielson:2022sga} was mainly carried out within the framework of quantum field theory in curved spacetime. More recently, it has been shown that these effects may also admit a local description \cite{Danielson:2024yru}, in which the decoherence rate can be related directly to the local Wightman function of the environmental quantum field. An important feature of quantum field theory in curved spacetime is that the notion of particles and vacuum states is observer dependent \cite{Birrell:1982ix,Wald:1995yp}. In curved spacetime, the absence of a unique global definition of positive-frequency modes renders the particle content of quantum fields ambiguous. In particular, black hole spacetimes admit three inequivalent vacuum states, namely the Boulware \cite{Boulware:1974dm}, Unruh \cite{Unruh:1976db}, and Hartle-Hawking vacua \cite{Hartle:1976tp}. The Boulware vacuum corresponds to the absence of particles with respect to Schwarzschild time and reduces to Minkovski vacuum at infinity. The Hartle-Hawking vacuum describes an eternal black hole in thermal equilibrium with thermal radiation both entering from infinity and emerging from the horizon. The Unruh vacuum models an evaporating black hole formed by gravitational collapse, where no incoming radiation comes from infinity while Hawking radiation is emitted from the horizon.

In this work, we investigate the decoherence of a static particle in a spatial superposition near the Schwarzschild and Reissner-Nordstr\"{o}m black holes. Based on the observation that a classical source linearly coupled to a quantum field generates a coherent state of the field \cite{Glauber:1963tx,2012JPhA...45x4002S,Zhang:1990fy}, we reformulate the decoherence process through coherent state generation of field excitations. This mechanism is closely analogous to the generation of coherent radiation by a classical current distribution in quantum optics \cite{Scully:1997hcl}. When a classical source couples linearly to a quantum field, the vacuum state evolves into a coherent state. Physically, the classical source acts as a coherent drive that continuously displaces the field modes, analogous to how a classical current drives the electromagnetic field in a laser to produce a coherent state of photons. Consequently, different classical source configurations generate different coherent radiation states, and the overlap between these coherent states naturally determines the decoherence of a quantum superposition of sources. By evaluating the overlap between field coherent states associated with different branches of the superposition, we derive a local expression for the decoherence functional in terms of the Wightman two-point function of the scalar field. In the long-time limit, the decoherence dynamics can be characterized by a local decoherence rate governed by the low-frequency behavior of field correlations.

We then apply the formalism to Schwarzschild and Reissner-Nordstr\"{o}m spacetimes and investigate the decoherence behavior in the Boulware, Unruh, and Hartle-Hawking vacua. Using low-frequency approximations of scalar field mode solutions based on the asymptotic matching method \cite{Das:1996we,Gubser:1996xe,Kanti:2002nr}, we obtain analytic expressions for the decoherence rate and analyze how the quantum state of the black hole environment influences quantum coherence. Our analysis may be viewed as a local quantum field theoretic realization of the DSW gedankenexperiment and provides a local description of environment induced decoherence in black hole spacetimes.

The remainder of this paper is organized as follows. In Sec.~\ref{sec:gsw}, we briefly review the basic set-up of the GSW gedankenexperiment. In Sec.~\ref{sec:dec_coh}, we discuss the coherent state formalism for environment induced decoherence and derive a local expression for the decoherence functional in terms of the field correlation function. Then by taking the long time limit, the decoherence rate is related to the zero-frequency limit of the Wightman function. In Sec.~\ref{sec:schwz}, we apply the formalism to Schwarzschild black holes, where we compute the scalar field Wightman functions in different vacua and analytically derive the corresponding decoherence rates. In Sec.~\ref{sec:rn}, we extend the analysis to Reissner-Nordström black holes. Finally, Sec.~\ref{sec:con_dis} is devoted to conclusions and discussions.

\section{GSW Gedankenexperiment }
\label{sec:gsw}

In this section, we briefly review the basic setup of the GSW gedankenexperiment \cite{Danielson:2022tdw,Danielson:2022sga}. Consider a charged massive particle held at a fixed position outside a static black hole. Suppose that the particle is initially prepared in the spin state polarized along the positive $x$ direction. The particle is then sent through a Stern-Gerlach apparatus oriented along the $z$ axis, which spatially separates the two spin components. After passing through the apparatus, the particle is prepared in the entangled spin-position superposition
\begin{eqnarray}\label{superposition_state}
    |\Psi\rangle
    =
    \frac{1}{\sqrt{2}}
    \left(
    |\psi_L,\uparrow\rangle
    +
    |\psi_R,\downarrow\rangle
    \right),
\end{eqnarray}
where $|\psi_L\rangle$ and $|\psi_R\rangle$ denote two normalized spatially separated wave packets, while $|\uparrow\rangle$ and $|\downarrow\rangle$ represent spin eigenstates along the $z$ direction. It is assumed that the two wave packets are sufficiently narrow and spatially separated so that they do not overlap. The superposition is maintained for a sufficiently long duration $T$, after which the two branches are recombined by a reversed Stern-Gerlach apparatus. In principle, one may verify whether the initial state is recovered by measuring the spin along the $x$ direction. If coherence is perfectly preserved during the holding time, the particle should always return to its initial spin state. It is also assumed that the splitting and recombination processes are performed adiabatically. Under this assumption, decoherence during the separation and recombination processes can be neglected \cite{Danielson:2022tdw}.

At first sight, one may expect the coherence of the superposition to remain intact, provided that the particle source is sufficiently isolated from external disturbances. However, the electromagnetic or gravitational fields coupled to the particle source should also be treated as part of the total system. The two branches of the spatial superposition induce different field configurations. As long as the splitting and recombination processes are performed adiabatically, the emission of entangling radiation to null infinity can be made arbitrarily small \cite{Danielson:2022tdw}. Under this condition, it has been argued that coherence can in principle be preserved in Minkowski spacetime  \cite{Belenchia:2018szb,Danielson:2021egj}. The situation becomes qualitatively different in black hole spacetimes. Even in the adiabatic limit, low-frequency electromagnetic or gravitational radiation can be absorbed by the black hole horizon. Due to the causal structure of the horizon, this radiation becomes inaccessible to exterior observers and effectively carries away ``which-way'' information of the particle, analogous to an observer hidden behind the horizon performing a measurement. Consequently, the coherence of the spatial superposition is inevitably degraded in black hole backgrounds.

One may wonder whether this type of decoherence can be avoided if the charged massive particle freely falls into the black hole. Although a freely falling observer locally experiences an approximately inertial frame, a charged particle in Schwarzschild spacetime generally still radiates both into the black hole and to infinity \cite{Scully:2017utk,Shallue:2025zto}. The electromagnetic or gravitational fields associated with different branches of the quantum superposition may remain distinguishable and carry ``which-way'' information, thereby inducing decoherence. Unlike the stationary setup considered in the original DSW analysis, where the charge follows a timelike Killing trajectory and minimizes radiation, the decoherence of a freely falling superposition is not expected to obey a universal bound \footnote{We thank Dr.~Gautam Satishchandran for helpful discussions on this point through private communication.}.

\section{Decoherence of spatial superposition state from the coherent emission}
\label{sec:dec_coh}

In this section, we develop a general formalism for the decoherence of a spatial quantum superposition induced by the coupling between a particle and a scalar field in a static black hole spacetime. Our derivation is based on the fact that a localized classical source linearly coupled to a quantum field generates a coherent state of the field.

\subsection{Coherent scalar radiation from classical source} 

Let us consider the globally hyperbolic static, spherically symmetric spacetime described by the metric 
\begin{equation}\label{metric}
    ds^2=-f(r)dt^2+\frac{dr^2}{f(r)}+r^2d\Omega^2,
\end{equation}
where $d\Omega^2$ denotes the metric on the 2-sphere. For the Schwarzschild black holes, the metric function $f(r)$ is given by  
\begin{eqnarray}
    f(r)=1-\frac{2M}{r}\;,
    \end{eqnarray}
where $M$ is the mass of the black hole. However, the following discussion in this section is independent of the specific form of the metric function. 

We will treat a point particle as a classical source interacting with a real scalar field $\Phi$ in the background of the spacetime described by Eq.\eqref{metric}. The covariant interaction between a point particle and a scalar field is given by 
\begin{equation}
S_{\text{int}} = \lambda \int d\tau \Phi(x(\tau)),
\end{equation}
where $\lambda$ is the coupling constant and $x(\tau)$ is the worldline of the point particle parameterized by the propertime $\tau$. This expression is manifestly covariant and describes a point particle following a worldline $x^\mu(\tau)$ interacting locally with the scalar field. It can be rewritten as 
\begin{equation}
S_{\text{int}} = \int d^4x \sqrt{-g} J(x)\Phi(x),
\end{equation}
where the source is given by
\begin{equation}
J(x) = \lambda \int d\tau \frac{\delta^{(4)}(x - x(\tau))}{\sqrt{-g(x)}}.
\end{equation}
Here $\delta^{(4)}(x - x(\tau))$ is conventional the coordinate delta function.

The interaction Hamiltonian can be defined on arbitrary Cauchy surface with the induced metric $h_{\mu\nu}$ and the volume element $\sqrt{h}d^3 x$. For the static black holes considered here, we take the constant $t$ hypersurface as the Cauchy surface. For the point particle interacting with the scalar field, the interaction Hamiltonian is defined as 
\begin{eqnarray}
    H_{I}(t) = -\int d^3x \sqrt{-g}  J(x)\Phi(x).
\end{eqnarray}
Substituting the point particle source and performing the spatial integration yields
\begin{equation}
H_{I}(t)=
-\lambda \int d\tau \delta(t - t(\tau)) \Phi(x(\tau)).
\end{equation}

By using the property of $\delta$-function, 
\begin{eqnarray}
    \delta(t - t(\tau)) = \frac{\delta(\tau - \tau(t))}{dt/d\tau},
\end{eqnarray}
one can obtain
\begin{equation}
H_{I}(t) = -\lambda \frac{d\tau}{dt} \Phi(x(t)).
\end{equation}
In the interaction picture, the time evolution operator is given by 
\begin{equation}
U_I(t,t_0) = \mathcal{T} \exp\left(-i \int_{t_0}^{t} dt' H_I(t')\right),
\end{equation}
where $\mathcal{T}$ denotes the time-ordered product. For the present case, it becomes 
\begin{equation}\label{U_opr}
U_I(t,t_0)  =
\mathcal{T}
\exp\left(
i \lambda \int d\tau \Phi(x(\tau))
\right).
\end{equation}
This form is fully covariant and shows that the evolution is driven by the field evaluated along the worldline of the point particle.

For the quantized scalar field operator $\Phi(x)$, it can be expanded as \cite{Birrell:1982ix}
\begin{equation}\label{scalar_exp}
\Phi(x) = \sum_k \left(a_k u_k(x) + a_k^\dagger u_k^*(x)\right),
\end{equation}
where $u_k(x)$ denotes the mode function and the operators $a_k$ and $a_k^\dagger$ are the annihilation and creation operators satisfying the conventional commutation relation 
\begin{eqnarray}
    [a_k,a_{k'}^\dagger]=\delta_{k,k'}.
\end{eqnarray}
The annihilation operator $a_k$ can be used to define the vacuum state $|0\rangle$ of the scalar field  
\begin{eqnarray}
    a_k |0\rangle =0.
\end{eqnarray}
It should be noted that the index $k$ represents a set of quantities necessary to label the mode functions in general.

Assume that at the initial time the state of the scalar field is in the vacuum state. In the interacting picture, the quantum state at later time is given in terms of the time evolution operator  
\begin{eqnarray}
    |\Psi(t)\rangle= U_I(t,t_0) |0\rangle\;. 
\end{eqnarray}

By substituting the expansion of the scalar field operator Eq.\eqref{scalar_exp} into the time evolution operator Eq.\eqref{U_opr} and defining the following coefficient as 
\begin{equation}
\alpha_k = i\lambda \int d\tau u_k^*(x(\tau)),
\end{equation}
the quantum state of the scalar field at the time $t$ is given by 
\begin{equation}\label{Coherent_Sta_D}
|\Psi(t)\rangle=
\exp\left(\sum_k \left(\alpha_k a_k^\dagger-\alpha_k^* a_k\right)\right)
|0\rangle,
\end{equation}
where an overall phase factor has been ignored. It is manifest that the quantum state at the time $t$ under the interacting of point particle source with the scalar field is a coherent state.

The displacement operator can be defined as \cite{Glauber:1963tx}
\begin{eqnarray}
    D(\alpha)=\exp\left(\sum_k \left(\alpha_k a_k^\dagger-\alpha_k^* a_k\right)\right).
\end{eqnarray}
Eq.\eqref{Coherent_Sta_D} shows that the evolved state $|\Psi(t)\rangle$ is a coherent state obtained by applying the displacement operator on the vacuum state. By using the Baker-Hausdorff formula, the evolved state $|\Psi(t)\rangle$ can be rewritten as 
\begin{equation}\label{Coherent_Sta}
|\Psi(t)\rangle=
\exp\left(-\frac{1}{2}\sum_k |\alpha_k|^2\right)
\exp\left(\sum_k \alpha_k a_k^\dagger\right)\exp\left(\sum_k -\alpha_k^* a_k\right) |0\rangle.
\end{equation}

Physically, this result reflects the correspondence between classical field configurations and coherent states. A classical source generates a classical wave configuration, whose quantum counterpart is naturally described by a coherent state of the field. This mechanism is closely analogous to coherent state generation in quantum optics, where a classical current linearly coupled to the quantized electromagnetic field produces a coherent radiation state \cite{Scully:1997hcl}. In the present setup, the localized classical source coupled to the scalar field similarly generates a coherent state, representing the quantum counterpart of the classical field configuration.

\subsection{Decoherence due to scalar radiation}

Now we consider the DSW gedankenexperiment. After the charged particle passing through the Stern-Gerlach apparatus, the particle is in a spatially separated superposition state, which effectively results in two possible evolutions of the radiation field. The state of the total particle-radiation system is then given by the following form 
\begin{eqnarray}\label{superposition_state_total}
    |\Psi\rangle= \frac{1}{\sqrt{2}}\left(|\psi_L,\uparrow\rangle\otimes |\Psi_L\rangle+|\psi_R,\downarrow\rangle\otimes |\Psi_R\rangle\right)\;,
\end{eqnarray}
where $|\Psi_L\rangle$ and $|\Psi_R\rangle$ denote the states of the scalar field radiation associated with the particle states $|\psi_L\rangle$ and $|\psi_R\rangle$. Comparing with the state given by Eq.~\eqref{superposition_state}, Eq.~\eqref{superposition_state_total} explicitly incorporates the scalar field radiation as part of the total quantum system. 

To quantify the decoherence induced by the scalar radiation, following DSW \cite{Danielson:2022tdw}, we define the decoherence measure of the superposition state as
\begin{eqnarray}
    \mathcal{D}=1-\left|\langle \Psi_L|\Psi_R\rangle\right|\;.
\end{eqnarray}
This quantity characterizes the distinguishability between the scalar field states associated with the two branches of the quantum superposition.

According to the general theory of coherent states \cite{Glauber:1963tx,Zhang:1990fy,2012JPhA...45x4002S}, we have shown that the radiation states $|\Psi_L\rangle$ and $|\Psi_R\rangle$ generated by the left and right sources are precisely the multi-mode coherent states explicitly given by Eq.\eqref{Coherent_Sta}. In terms of Eq.\eqref{Coherent_Sta_D}, they can also be written as 
\begin{eqnarray}
    |\Psi_{L/R}\rangle=
D(\alpha^{L/R})
|0\rangle,
\end{eqnarray}
where the displacement operators are given by
\begin{eqnarray}
     D(\alpha^{L/R})=\exp\left(\sum_k \left(\alpha_k^{L/R} a_k^\dagger-\alpha_k^{L/R*} a_k\right)\right),
\end{eqnarray}
with the displacement parameter 
\begin{equation}\label{alpha_def}
\alpha_k^{L/R} = i\lambda \int d\tau_{L/R} u_k^*(x_{L/R}(\tau_{L/R})).
\end{equation}
Here, $x_{L/R}(\tau_{L/R})$ denotes the spacetime trajectories of the two components of the spatial superposition, where $\tau_{L/R}$ is the proper time along the possible left/right trajectory.

We now calculate their overlap $\langle \Psi_L|\Psi_R\rangle$. Using the unitarity of the displacement operator, i.e. $D^\dagger(\alpha) = D(-\alpha)$, we write
\begin{eqnarray}
    \langle \Psi_L | \Psi_R \rangle = \langle 0 | D^\dagger(\alpha^L) D(\alpha^R) | 0 \rangle=\langle 0 | D(-\alpha^L) D(\alpha^R)| 0 \rangle.
\end{eqnarray}
The product of two displacement operators can be combined as
\begin{eqnarray}
D(-\alpha^L) D(\alpha^R) = e^{i\theta}  D(\beta)= e^{i\theta}\exp\left(\sum_k \left(\beta_k a_k^\dagger-\beta_k^* a_k\right)\right),
\end{eqnarray}
where $\beta = \alpha^{R} - \alpha^{L}$ and $\theta$ is a irrelevant phase factor that arises from the commutation of the exponents. Explicitly, $\beta_k$ is given by 
\begin{eqnarray}
    \beta_k=\alpha_k^{R} - \alpha_k^{L}=
    i\lambda \int d\tau \left(u_k^*(x_{R}(\tau))-u_k^*(x_{L}(\tau))\right).
\end{eqnarray}

Consequently, the overlap can be written as 
\begin{eqnarray}
\left|\langle \Psi_L | \Psi_R \rangle\right| = \langle 0 | D(\beta) | 0 \rangle.
\end{eqnarray}
For a single-mode displacement operator $D(\beta_k) = e^{\beta_k a^\dagger - \beta_k^* a}$, it is wellknown that the coherent state can be expanded as the superposition of the eigenstates 
\begin{eqnarray}
D(\beta_k)|0\rangle =  e^{-|\beta_k|^2/2} \sum_{n=0}^{\infty} \frac{\beta_k^n}{\sqrt{n!}} |n\rangle_k.
\end{eqnarray}
For multiple-mode coherent state, we have
\begin{eqnarray}
\langle 0 | D(\beta) | 0 \rangle = \prod_k \langle 0 | D(\beta_k) | 0 \rangle = \prod_k e^{-|\beta_k|^2/2} = \exp\!\left( -\frac12 \sum_k |\beta_k|^2 \right).
\end{eqnarray}
Substituting $\beta_k = \alpha_k^{R} - \alpha_k^{L}$, we have
\begin{eqnarray}\label{OverL_alpha}
\left|\langle \Psi_L | \Psi_R \rangle\right| = \exp\!\left( -\frac{1}{2} \sum_k \left|\alpha_k^{R} - \alpha_k^{L}\right|^2 \right).
\end{eqnarray}

Following DSW, the decoherence measure of the particles state is written as 
\begin{eqnarray}\label{D_N}
    \mathcal{D}=1-\exp\left[-\frac{1}{2} \langle N\rangle \right]\;,
\end{eqnarray}
where $\langle N\rangle$ is referred as expected number of entangling particles. If $\langle N\rangle$ is significantly larger than $1$, the superposition state \eqref{superposition_state_total} will be completely decohered. Combining Eq.\eqref{OverL_alpha} and Eq.\eqref{D_N}, the entangling particle number is then given by 
\begin{eqnarray}\label{N_exp}
  \langle N\rangle= \sum_k \left|\alpha_k^{R} - \alpha_k^{L}\right|^2 \;.
\end{eqnarray}
This is an intermediate result, which gives the relation between the entanglement particle number and the positive-frequency mode functions of the scalar field in terms of the definition of the displacement parameter $\alpha_k$. 

\subsection{The local expression of the decoherence rate}

Suppose that the two branches of the spatial superposition are placed at radial coordinates \(r_0-\frac{1}{2}\sqrt{f}d(\tau)\) and \(r_0 +\frac{1}{2}\sqrt{f} d(\tau)\) with \(d(\tau) \ll r_0\) being the proper distance between the two components. The difference between the two displacement parameters in the two branches can be approximated by 
\begin{eqnarray}
    \alpha_k^{R} - \alpha_k^{L} &=& i\lambda \int d\tau \; \left( u_k^*(x_R(\tau)) - u_k^*(x_L(\tau)) \right)\nonumber\\
    &=& i\lambda  \int d\tau\; d(\tau) s^\mu \nabla_\mu  u_k^*(t,x_0)\;,
\end{eqnarray}
where we have ignored the difference of the proper times in the left and the right superposition trajectories. Here $s^{\mu}$ is the unit tangent vector to the geodesic segment in a fixed $t$ hypersurface that connects the centers of mass of the two components, and the mode function $u_k(x)$ can be evaluated at the Alice's lab at time $t$ with $x_0=(r_0,\theta_0,\phi_0)$ being the spatial location.

Substituting this expression into Eq.\eqref{N_exp}, we can obtain 
\begin{eqnarray}
    \langle N\rangle=\lambda^2 \int d\tau d\tau' 
    d(\tau) d(\tau') \sum_k s^\mu \nabla_\mu  u_k^*(\tau,x_0) s^\nu \nabla_\nu  u_k(\tau',x_0).
\end{eqnarray}
Note that the integral is performed on the proper time and $d(\tau)$ is the proper distance between the two components. The expression for the entangling particle number is clearly coordinate invariant. By using the relation between the proper time and the coordinate time, we can rewrite the above integral as 
\begin{eqnarray}
    \langle N\rangle=\lambda^2 \int dt dt' f(r_0)
    d(t) d(t') \sum_k s^\mu \nabla_\mu  u_k^*(t,x_0) s^\nu \nabla_\nu  u_k(t',x_0).
\end{eqnarray}

If the separation of the two component is along the radial direction, the unit vector $s^\mu$ has only one component $s^r=\sqrt{f(r_0)}$. Then the integral can be further given by 
\begin{eqnarray}
    \langle N\rangle=\lambda^2 \int dt dt' f^2(r_0)
    d(t) d(t') \sum_k \partial_r  u_k^*(t,x_0) \partial_r u_k(t',x_0).
\end{eqnarray}
In this expression, $\langle N\rangle$ is expressed in terms of the positive-frequency mode functions $u_k(x)$. We now consider the relation between the entangling particle number and the correlation function of the scalar field. 

Recall that the positive-frequency Wightman function is defined as \cite{Birrell:1982ix}
\begin{eqnarray}
G_+(x,x') = \langle 0|\Phi(x)\Phi(x')|0\rangle 
\end{eqnarray}
By substituting the mode expansion of the scalar field operator into the above definition and using the definition of the vacuum state, one can obtain 
\begin{eqnarray}
   G_+(x,x') = \sum_k u_k(x) u_k^*(x'). 
\end{eqnarray}
This can be treated as the completeness relation of the mode function of the scalar field. Using the expression, the entangling particle number can be written as 
\begin{eqnarray}
   \langle N\rangle=\lambda^2 f^2(r_0) \int dt dt' 
    d(t) d(t') \left.\langle 0|\partial_r\Phi(x)\partial_{r'}\Phi(x')|0\rangle\right|_{r=r'=r_0} .
\end{eqnarray}

This expression for the entangling particle number is consistent with the one previously obtained in \cite{Li:2024lfv}. However, the method used here is slightly different from the method used in \cite{Danielson:2024yru}. In the following sections, we will employ this expression to calculate the decoherence of the spacial superposition states in the Schwarzschild black hole and the Reinsser-Nordstr\"{o}m black hole.

\subsection{Decoherence at the Long-time limit}

It is clear that the entanglement particle number depends on the choice of the separation function $d(t)$. Nevertheless, different choices of $d(t)$ generally lead to qualitatively similar results \cite{Fahn:2025fxl,Fahn:2025jxh}. For simplicity, we consider a rectangular separation profile, as in the original DSW analysis \cite{Danielson:2024yru}, where the superposition is maintained at a fixed separation for a finite duration.

We assume that $d(t)$ is a function as 
\begin{eqnarray}
    d(t)=\begin{cases}
d, & |t|<T/2\;,\\
0, & t<-T/2-T_1 \;\;\textrm{or}\;\; t>T/2+T_2\;,
\end{cases}
\end{eqnarray}
where $T$ is the coordinate time that the experimenter holds the superposition state and $T_1$ and $T_2$ are the times that used to separate and recombine the superposition. One can roughly approximate the function $d(t)$ as a rectangular wave function to estimate the entangling particle number. 

By using the Fourier transform of the separation function 
\begin{eqnarray}
    d(t)=\int \frac{d\omega}{2\pi} \tilde{d}(\omega) e^{-i\omega t}\;,
\end{eqnarray}
one can get 
\begin{eqnarray}
    \langle N\rangle &=&\lambda^2 f^2(r_0) \int \frac{d\omega}{2\pi} \frac{d\omega'}{2\pi} \tilde{d}^*(\omega) \tilde{d}(\omega')\nonumber\\
    &&\times\int dt dt' e^{i(\omega t-\omega' t')}\left.\langle 0|\partial_r\Phi(x)\partial_{r'}\Phi(x')|0\rangle\right|_{r=r'=r_0} .
\end{eqnarray}

In general, for the static spacetime, the correlation functions are time translation invariant and depend only on the time difference $t-t'$. Therefore, after changing the time variable to $\Delta t=t-t'$, we can get 
\begin{eqnarray}
     \langle N\rangle &=&\lambda^2 f^2(r_0) \int \frac{d\omega}{2\pi} \frac{d\omega'}{2\pi} \tilde{d}^*(\omega) \tilde{d}(\omega')\int dt' e^{i(\omega-\omega')t'} \mathcal{F}(\omega)\;,
\end{eqnarray}
where $\mathcal{F}(\omega)$ is defined as the Fourier transformation of the corelation function 
\begin{eqnarray}\label{Kernel_G}
   \mathcal{F}(\omega)=\int d(\Delta t)  e^{i\omega \Delta t}\left.\langle 0|\partial_r\Phi(x)\partial_{r'}\Phi(x')|0\rangle\right|_{r=r'=r_0}.
\end{eqnarray}
By evaluating the integral further, we can get 
\begin{eqnarray}
     \langle N\rangle &=&\lambda^2 f^2(r_0) \int \frac{d\omega}{2\pi} \frac{d\omega'}{2\pi} \tilde{d}^*(\omega) \tilde{d}(\omega') 2\pi\delta{(\omega-\omega')} \mathcal{F}(\omega)\nonumber\\
     &=&\lambda^2 f^2(r_0) \int \frac{d\omega}{2\pi}  \tilde{d}^*(\omega) \tilde{d}(\omega) \mathcal{F}(\omega)
     \nonumber\\
      &\approx&\lambda^2 f^2(r_0) \mathcal{F}(\omega=0) \int \frac{d\omega}{2\pi}  \tilde{d}^*(\omega) \tilde{d}(\omega) \nonumber\\
      &=&\lambda^2 f^2(r_0) \mathcal{F}(\omega=0) \int
     dt d^2(t)\nonumber\\
    &=&\lambda^2 f^2(r_0) \mathcal{F}(\omega=0) d^2 T \;.
\end{eqnarray}

Since $d(t)$ can be approximated as a rectangular wave function, for large $T$, $\tilde{d}(\omega)$ is highly bandlimited near $\omega\sim 0$. We have used this approximation in the above derivation. Note that we have also used the Parseval theorem for the Fourier transformation in the above derivation. This result indicates that, for large \(T\), the entangling particle number $\langle N\rangle$ is proportional to the separation time $T$. Then we can define the decoherence rate as 
\begin{eqnarray}\label{Gamma_F}
    \Gamma=\frac{d \langle N\rangle }{dT}= \lambda^2 d^2 f^2(r_0) \mathcal{F}(\omega=0) \approx \lambda^2 d^2 \mathcal{F}(\omega=0)\;,
\end{eqnarray}
where we have used the approximation that the lab is located far from the black hole and $f(r_0)\approx 1$. The final task is then to calculate the Fourier transformation of the correlation function for different static black holes.

\section{Local calculation of decoherence in Schwarzschild black holes} 
\label{sec:schwz}

In this section, we present the detailed calculation of the decoherence for a quantum spatial superposition state in the Schwarzschild black hole. The essential is to compute the two piont correlation function of the scalar field in the Schwarzschild black hole.

\subsection{Wightman function of scalar field in Schwarzschild black holes}

The dynamics of the massless scalar field is governed by the Klein-Gordon equation 
\begin{eqnarray}
    \nabla_{\mu}\nabla^{\mu}\Phi(x)=0\;,
\end{eqnarray}
where we have set the mass of the scalar field to be vanishing. 

We consider the positive-frequency mode solutions of the Klein-Gordon equation in the Schwarzschild black hole 
\begin{eqnarray}\label{variable_separating}
    u_{\omega lm}(x)=\frac{1}{\sqrt{4\pi\omega}}\frac{R_{\omega l}(r)}{r} e^{-i\omega t}  Y_{lm}(\theta,\phi)\;, 
\end{eqnarray}
where $\omega>0$ and $Y_{lm}(\theta,\phi)$ is the spherical harmonic function. By substituting the expression for the mode function into the Klein-Gordon equation, one can get the radial equation for the radial function $R_{\omega l}(r)$ as 
\begin{eqnarray}\label{radial_eq}
    \frac{d^2 R_{\omega l}}{dr^{*2}}+\left[\omega^2-V(r)\right]R_{\omega l}(r)=0\;,
\end{eqnarray}
with the effective potential 
\begin{eqnarray}
    V(r)=\left(1-\frac{2M}{r}\right)\left(\frac{l(l+1)}{r^2}+\frac{2M}{r^3}\right)\;,
\end{eqnarray}
and the tortoise coordinate 
\begin{eqnarray}
    r^*=r+2M\ln\left(\frac{r}{2M}-1\right). 
\end{eqnarray}

In the asymptotic limit $r\to\infty$, the radial equation admits solutions of the form $R_{\omega l}(r)\sim e^{\pm i\omega r_*}$. The mode solutions whose leading-order behavior at infinity is purely outgoing wave, $R_{\omega l}(r)\sim e^{+ i\omega r_*}$ are referred to as the up-modes. Despite their simple outgoing form at spatial infinity, these modes become a linear superposition of ingoing and outgoing components near the horizon due to scattering by the effective potential. Therefore, the up-modes are defined as the modes that originate from past event horizon $\mathcal{H}^{-}$. The boundary conditions satisfied by the up-modes are given by \cite{Fabbri:1975sa,Candelas:1980zt}
\begin{align}\label{up_bc}
    R^{\mathrm{up}}_{\omega l}(r) \sim \begin{cases}
       e^{+i\omega r^*}+A^{\mathrm{up}}_{\omega l} e^{-i\omega r^*} , &r\rightarrow 2M \\
       B^{\mathrm{up}}_{\omega l} e^{+i\omega r^*} , &r\rightarrow \infty
    \end{cases}
\end{align}

Conversely, the in-modes are defined to be purely ingoing at the horizon, $R_{\omega l}(r)\sim e^{-i\omega r_*}$, but are scattered into a linear combination of ingoing and outgoing waves at infinity. Therefore, the in-modes are defined as the modes that originate from past null infinity $\mathcal{J}^{-}$. The boundary conditions satisfied by the in-modes are given by \cite{Fabbri:1975sa,Candelas:1980zt}
\begin{align}\label{in_bc}
    R^{\mathrm{in}}_{\omega l}(r) \sim \begin{cases}
        A^{\mathrm{in}}_{\omega l} e^{-i\omega r^*}, &r\rightarrow 2M \\
        e^{-i\omega r^*}+B^{\mathrm{in}}_{\omega l} e^{+i\omega r^*}, &r\rightarrow \infty
    \end{cases}
\end{align}

For $\omega>0$, $\{u_{\omega lm}^{\mathrm{in}}(x),u_{\omega lm}^{\mathrm{up}}(x)\}$ are the positive frequency modes, and its conjugate modes $\{u_{\omega lm}^{\mathrm{in}*}(x),u_{\omega lm}^{\mathrm{up*}}(x)\}$ are the corresponding negative frequency modes. These modes form a complete basis that satisfies the conventional orthonormality relations. The scalar field can be expanded by using this set of mode functions. Then, the quantized scalar field, which is quantized by replacing the expansion coefficients by the annihilation and the creation operators, can be obtained.

However, we should distinguish three different vacua. According to quantum field theory on the curved spacetime background, the notion of a “vacuum” is not unique. For a static, spherically symmetric black hole like Schwarzschild black hole, there are three kinds of vacua, i.e. the Boulware vacuum \cite{Boulware:1974dm}, the Unruh vacuum \cite{Unruh:1976db} and the Hartle-Hawking vacuum \cite{Hartle:1976tp}.

For the Boulware vacuum, the quantum field can be expanded by using the ingoing and the outgoing Boulware modes $\{u_{\omega lm}^{\mathrm{in}}(x),u_{\omega lm}^{\mathrm{up}}(x),u_{\omega lm}^{\mathrm{in}*}(x),u_{\omega lm}^{\mathrm{up*}}(x)\}$ that are defined with respect to the Schwarzschild coordinate time $t$. It is given by 
\begin{eqnarray}
    \Phi(x)=\sum_{l=0}^{+\infty}\sum_{m=-l}^{l}\int_0^{+\infty} d\omega \left(
    \hat{b}_{\omega l m}^{\mathrm{up}}u_{\omega l m}^{\mathrm{up}}+\hat{b}_{\omega l m}^{\mathrm{in}}u_{\omega l m}^{\mathrm{in}}+\hat{b}_{\omega l m}^{\mathrm{up}\dagger}u_{\omega l m}^{\mathrm{up}*}+\hat{b}_{\omega l m}^{\mathrm{in}\dagger}u_{\omega l m}^{\mathrm{in}*}\right)\;.
\end{eqnarray}
where the $\hat{b}$ and $\hat{b}^\dagger$ operators are respectively the annihilation and creation operators for the $u$ modes. The Boulware vacuum is then defined as 
\begin{eqnarray}
    \hat{b}^{\mathrm{up}}|0\rangle_B=\hat{b}^{\mathrm{in}}|0\rangle_B=0\;.
\end{eqnarray}
It can seen that the Boulware vacuum is defined by the positive frequency modes with respect to the Schwarzschild Killing time. It is analogous to the Rindler vacuum and is regular at spatial infinity, but becomes singular at the horizon. It corresponds to a static black hole with no thermal radiation.

For the Unruh and the Hartle-Hawking vacuum, the Wightman functions can be obtained by expanding the quantum field in terms of modes that are positive frequency with respect to the affine parameters on the past and future horizons. These modes are given by the following mode transformation \cite{Unruh:1976db,Christensen:1977jc}
\begin{align}
w^{\mathrm{in}}_{\omega\ell m} &= \frac{1}{\sqrt{2\sinh(4\pi M\omega)}} \bigl( e^{2\pi M\omega} u^{\mathrm{in}}_{\omega\ell m} + e^{-2\pi M\omega} v^{\mathrm{in}*}_{\omega\ell m} \bigr), \\
\bar{w}^{\mathrm{in}}_{\omega\ell m} &= \frac{1}{\sqrt{2\sinh(4\pi M\omega)}} \bigl( e^{-2\pi M\omega} u^{\mathrm{in}*}_{\omega\ell m} + e^{2\pi M\omega} v^{\mathrm{in}}_{\omega\ell m} \bigr), \\
w^{\mathrm{up}}_{\omega\ell m} &= \frac{1}{\sqrt{2\sinh(4\pi M\omega)}} \bigl( e^{2\pi M\omega} u^{\mathrm{up}}_{\omega\ell m} + e^{-2\pi M\omega} v^{\mathrm{up}*}_{\omega\ell m} \bigr), \\
\bar{w}^{\mathrm{up}}_{\omega\ell m} &= \frac{1}{\sqrt{2\sinh(4\pi M\omega)}} \bigl( e^{-2\pi M\omega} u^{\mathrm{up}*}_{\omega\ell m} + e^{2\pi M\omega} v^{\mathrm{up}}_{\omega\ell m} \bigr).
\end{align}
where the $v$ modes are functions analogous to the $u$ modes on the second exterior region of the Kruskal manifold. The modes are extended to the full Kruskal manifold by analytic continuation. Note that the $u$ functions are non-vanishing on the first exterior region of the Kruskal manifold and the $v$ functions are non-vanishing on the second exterior region of the Kruskal manifold. The mode transformation presented in the above equation mixes the ingoing Boulware modes with the outgoing Boulware modes.

For the Unruh vacuum, the quantum field can be expanded as 
\begin{eqnarray}
    \Phi(x)=\sum_{l=0}^{+\infty}\sum_{m=-l}^{l}\int_0^{+\infty} d\omega && \left(
    \hat{d}_{\omega l m}^{\mathrm{up}}w_{\omega l m}^{\mathrm{up}}+ \hat{\bar{d}}_{\omega l m}^{\mathrm{up}}\bar{w}_{\omega l m}^{\mathrm{up}}+\hat{b}_{\omega l m}^{\mathrm{in}}u_{\omega l m}^{\mathrm{in}}\right.
    \nonumber\\
    &&\left.+ \hat{d}_{\omega l m}^{\mathrm{up}\dagger}w_{\omega l m}^{\mathrm{up}*}+ \hat{\bar{d}}_{\omega l m}^{\mathrm{up}\dagger}\bar{w}_{\omega l m}^{\mathrm{up}*}+\hat{b}_{\omega l m}^{\mathrm{in}\dagger}u_{\omega l m}^{\mathrm{in}*}\right)\;.
\end{eqnarray}
where the $\hat{b}$ and $\hat{b}^\dagger$ operators are respectively the annihilation and creation operators for the $u$ modes. The Unruh vacuum is defined as 
\begin{eqnarray}
    \hat{d}^{\mathrm{up}}|0\rangle_U=\hat{\bar{d}}^{\mathrm{up}}|0\rangle_U=\hat{b}^{\mathrm{in}}|0\rangle_U=0\;.
\end{eqnarray}
It can be seen that the Unruh vacuum is defined by the positive frequency modes with respect to the Kruskal affine parameter on the future horizon and with respect to the usual Schwarzschild time at past null infinity. It is singular on the past horizon but regular on the future horizon. It describes that the state that a black hole formed by gravitational collapse. 

For the Hartle-Hawking vacuum, the quantum field can be expanded as 
\begin{eqnarray}
    \Phi(x)=\sum_{l=0}^{+\infty}\sum_{m=-l}^{l}\int_0^{+\infty} d\omega && \left(
    \hat{d}_{\omega l m}^{\mathrm{up}}w_{\omega l m}^{\mathrm{up}}+ \hat{\bar{d}}_{\omega l m}^{\mathrm{up}}\bar{w}_{\omega l m}^{\mathrm{up}}+\hat{d}_{\omega l m}^{\mathrm{in}}w_{\omega l m}^{\mathrm{in}}+ \hat{\bar{d}}_{\omega l m}^{\mathrm{in}}\bar{w}_{\omega l m}^{\mathrm{in}}\right.
    \nonumber\\
    &&\left.+ \hat{d}_{\omega l m}^{\mathrm{up}\dagger}w_{\omega l m}^{\mathrm{up}*}+ \hat{\bar{d}}_{\omega l m}^{\mathrm{up}\dagger}\bar{w}_{\omega l m}^{\mathrm{up}*}+\hat{d}_{\omega l m}^{\mathrm{in}\dagger}w_{\omega l m}^{\mathrm{in}*}+ \hat{\bar{d}}_{\omega l m}^{\mathrm{in}\dagger}\bar{w}_{\omega l m}^{\mathrm{in}*}\right)\;.
\end{eqnarray}
where the $\hat{b}$ and $\hat{b}^\dagger$ operators are respectively the annihilation and creation operators for the $u$ modes. The Hartle-Hawking vacuum is defined as 
\begin{eqnarray}
    \hat{d}^{\mathrm{up}}|0\rangle_H=\hat{\bar{d}}^{\mathrm{up}}|0\rangle_H= \hat{d}^{\mathrm{in}}|0\rangle_H=\hat{\bar{d}}^{\mathrm{in}}|0\rangle_H=0\;.
\end{eqnarray}
The Hartle‑Hawking vacuum is defined by the positive frequency modes with respect to the affine parameters on the past and future horizons. It is globally regular across both the past and future horizons and describes a black hole in thermal equilibrium with a heat bath at the Hawking temperature.

By using the expansion of the quantum field and the commutation relations between the annihilation and creation operators, one can obtain the Wightman functions in the exterior region for three kinds of vacua as \cite{Candelas:1980zt,Hodgkinson:2014iua,Ng:2014kha,Caribe:2023fhr}
\begin{align}
G_+(x,x') &\equiv \langle 0 | \Phi(x)\Phi(x') | 0 \rangle \\
&= \sum_{l=0}^{+\infty} \frac{(2l+1) P_l(\cos\gamma)}{16\pi^2rr'}  \int_{-\infty}^{+\infty} \frac{d\omega}{\omega} e^{-i\omega\Delta t} G_{\omega l}(r,r'),
\end{align}
where $\gamma$ is the angular separation between the two spacetime points $x$ and $x'$, $P_l(\cos\gamma)$ is the Legendre polynomial with the argument $\cos\gamma$, and the integral kernel $G_{\omega l}(r,r')$ depends on the quantum state of the field. The integral kernel takes the forms
\begin{eqnarray}
G_{\omega l}^B(r,r')&=&\Theta(\omega)\Bigl[ R_{\omega l}^{\mathrm{up}}(r)R_{\omega l}^{\mathrm{up}*}(r') + R_{\omega l}^{\mathrm{in}}(r)R_{\omega l}^{\mathrm{in}*}(r') \Bigr]\;,\\
G_{\omega l}^U(r,r')&=&\frac{1}{1-e^{-8\pi M \omega}}  R_{\omega l}^{\mathrm{up}}(r)R_{\omega l}^{\mathrm{up}*}(r') + \Theta(\omega) R_{\omega l}^{\mathrm{in}}(r)R_{\omega l}^{\mathrm{in}*}(r') \;,\\
G_{\omega l}^H(r,r')&=&\frac{1}{1-e^{-8\pi M \omega}}\Bigl[  R_{\omega l}^{\mathrm{up}}(r)R_{\omega l}^{\mathrm{up}*}(r') + R_{\omega l}^{\mathrm{in}}(r)R_{\omega l}^{\mathrm{in}*}(r')\Bigr]\;,
\end{eqnarray}
where $\Theta(\omega)$ are the Heaviside step function and the index $B,U$ and $H$ denote the Boulware vacuum, Unruh vacuum and Hartle-Hawking vacuum, respectively. 

By substituting the Wightman function into Eq.\eqref{Kernel_G}, one can get
\begin{eqnarray}
    \mathcal{F}(\omega)= \frac{1}{8\pi} \sum_{l=0}^{+\infty} \frac{(2l+1)}{\omega} \partial_r\partial_{r'}\left.\left(\frac{G_{\omega l(r,r')}}{rr'}\right)\right|_{r=r'=r_0}\;,
\end{eqnarray}
where we have used the fact that $P_l(\cos\gamma=1)$ for $x^i=x'^{i}$. According to Eq.\eqref{Gamma_F}, the zero-frequency limit of $\mathcal{F}(\omega)$ determines the decoherence rate.

\subsection{Analytical derivation of mode functions}

In order to obtain the decoherence rate $\Gamma$, we just need to calculate $\mathcal{F}(\omega=0)$, i.e. we just need to calculate the mode solutions of the scalar field at the low frequency. In the following, we will compute the approximated solutions of the scalar field modes in terms of the asymptotic matching method \cite{Das:1996we,Gubser:1996xe,Kanti:2002nr}.

We start with the radial equation \eqref{radial_eq}. In the near horizon region, by introducing the variable $z=1-\frac{2M}{r}$, the radial equation can be approximated as  
\begin{eqnarray}
    z(1-z)\frac{d^2 R}{dz^2}+(1-3z)\frac{dR}{dz}+\left[\sigma^2-\frac{l(l+1)}{1-z}-1\right]R=0\;,
\end{eqnarray}
where $\sigma=2M\omega$. This equation is only valid in the low frequency limit $M\omega\ll0$. By introducing an appropriate change of variables, this equation can be transformed into a hyper-geometric equation. 

Let
\begin{eqnarray}
    R(z)=z^{\alpha}(1-z)^\beta F(z)\;,
\end{eqnarray}
the equation can be further rewritten as 
\begin{eqnarray}\label{hyper_Geo_eq}
    z(1-z)F''(z)+\left[c-(1+a+b)\right]F'(z)-abF(z)\;,
\end{eqnarray}
where
\begin{eqnarray}
    c=1+2\alpha\;,\;\;\; a=b=1+\alpha+\beta\;,
\end{eqnarray}
with 
\begin{eqnarray}
    \alpha=-i\sigma\;,\;\;\;\beta=-\frac{1}{2}-\sqrt{\left(l+\frac{1}{2}\right)^2-\sigma^2}\;.
\end{eqnarray}
Eq.\eqref{hyper_Geo_eq} is the well known hyper-geometric equation. By using the general solution of the hyper-geometric equation, the solution to the radial equation can be written as 
\begin{eqnarray}
    R_{NH}&=&A_- z^\alpha (1-z)^\beta F(a,b,c;z)\nonumber\\
    &&+A_+ z^{-\alpha} (1-z)^\beta F(a-c+1,b-c+1,2-c; z)\;,
\end{eqnarray}
where $A_-$ and $A_+$ are arbitrary constants. The first term represents the ingoing solution and the second term represents the outgoing solution. 

We firstly consider the in-modes with the boundary conditions given by Eq.\eqref{in_bc}. For in-modes, there is no outgoing wave near the horizon. Then we can set $A_+=0$. The near horizon solution can be given by 
\begin{eqnarray}\label{NH_sol}
    R_{NH}=A_- z^\alpha (1-z)^\beta F(a,b,c;z)\;.
\end{eqnarray}

In the far-field region, where $r\gg M$, the radial equation \eqref{radial_eq} can be approximated as 
\begin{eqnarray}
    \frac{d^2 R}{dr^2}+\left[\omega^2-\frac{l(l+1)}{r^2}\right]R=0\;.
\end{eqnarray}
The substitution $R=\sqrt{r}\bar{R}$  brings the above equation into the form of Bessel equation for $\bar{R}$, which is given by 
\begin{eqnarray}
    \frac{d^2\bar{R}}{dr^2}+\frac{1}{r}\frac{d\bar{R}}{dr}+\left[\omega^2-\frac{\left(l+\frac{1}{2}\right)^2}{r^2}\right]\bar{R}=0\;.
\end{eqnarray}
Using the Bessel functions of the first kind, the general solution to the radial equation in the far field region can be given by 
\begin{eqnarray}\label{FF_sol}
    R_{FF}=B_1 \sqrt{r} J_{l+\frac{1}{2}}(\omega r)+B_2 \sqrt{r} Y_{l+\frac{1}{2}}(\omega r)\;.
\end{eqnarray}
where $J_{l+\frac{1}{2}}(\omega r)$ and $Y_{l+\frac{1}{2}}(\omega r)$ denote the spherical Bessel and Neumann functions.

When $r\rightarrow \infty$, the solution $R_{FF}$ can be further approximated as 
\begin{eqnarray}
    R_{FF}\rightarrow \frac{1}{\sqrt{2\pi \omega}} \left[(B_1+iB_2)e^{-i\left[\omega r-\frac{\pi}{2}(l+1)\right]}+(B_1-iB_2)e^{i\left[\omega r-\frac{\pi}{2}(l+1)\right]}\right]\;.
\end{eqnarray}
It is clear that the first term represents the ingoing wave and the second term represents the outgoing wave. To match the boundary conditions for the in-modes, we must have the constraint condition
\begin{eqnarray}\label{B_con_eq}
    \frac{1}{\sqrt{2\pi \omega}}(B_1+iB_2)e^{i\frac{\pi}{2}(l+1)}=1\;.
\end{eqnarray}

When the low frequency condition $\omega M\ll 1$ holds, there will be a large intermediate region between the near-horizon region and the far-field region. We assume that the Alice's lab is located in the intermediate region $M\ll r\ll\frac{1}{\omega}$. The approximated solution to the radial equation can be obtained by smoothly matching the near-horizon solution with the far-field solution. 

Before doing so, we first need to extrapolate the two solutions
towards the intermediate region. By using the transformation property of the hypergeometric function, the near-horizon solution \eqref{NH_sol} can be transformed into the form of 
\begin{eqnarray}
    R_{NH}&=&A_- z^\alpha (1-z)^\beta \left[
    \frac{\Gamma(c)\Gamma(c-a-b)}{\Gamma(c-a)\Gamma(c-b)}F(a,b,a+b-c+1;1-z)\right.\nonumber\\ &&\left. 
   +(1-z)^{c-a-b} \frac{\Gamma(c)\Gamma(a+b-c)}{\Gamma(a)\Gamma(b)}F(c-a,c-b,c-a-b+1;1-z)
    \right]\;.
\end{eqnarray}
For large $r$, i.e. $z\rightarrow 1$, the near-horizon solution \eqref{NH_sol} can be approximated as 
\begin{eqnarray}
    R_{NH}\sim A_1 \left(\frac{r}{2M}\right)^{l+1} +A_2 \left(\frac{r}{2M}\right)^{-l} \;,
\end{eqnarray}  
with    
\begin{eqnarray}    
   A_1= A_- \frac{\Gamma(c)\Gamma(c-a-b)}{\Gamma(c-a)\Gamma(c-b)}\;,\;\;
   A_2=A_-\frac{\Gamma(c)\Gamma(a+b-c)}{\Gamma(a)\Gamma(b)}\;.
\end{eqnarray}

For small $r$, the far-field solution \eqref{FF_sol} can be approximated as 
\begin{eqnarray}
    R_{FF}\sim \frac{B_1}{\Gamma(l+\frac{3}{2})}\left(\frac{\omega}{2}\right)^{l+\frac{1}{2}}r^{l+1}-\frac{B_2}{\pi}\Gamma(l+\frac{1}{2})\left(\frac{\omega}{2}\right)^{-l-\frac{1}{2}}r^{-l}\;.
\end{eqnarray}
The matching of the two solutions gives us the following conditions 
\begin{eqnarray}
    A_1 &=&\frac{B_1}{\Gamma(l+\frac{3}{2})} (M\omega)^{l+1} \sqrt{\frac{2}{\omega}}\;,\nonumber\\
    A_2 &=&-\frac{B_2}{\pi} \Gamma(l+\frac{1}{2}) (M\omega)^{-l} \sqrt{\frac{2}{\omega}}\;.
\end{eqnarray}
From these two matching consition, one can easily see that 
\begin{eqnarray}
    B_2\sim B_1 (M\omega)^{2l+1}\ll B_1\;.
\end{eqnarray}
In the low frequency region, we can take $B_2=0$. Then the constraint equation \eqref{B_con_eq} gives 
\begin{eqnarray}
    B_1=\sqrt{2\pi \omega} (-i)^{l+1}\;,
\end{eqnarray}
which in turn gives the approximated radial solution in the intermediate region as 
\begin{eqnarray}
   R_{\omega l}^{\mathrm{in}}(r)\approx \frac{i^{3l+3}2^{l+1}l!}{(2l+1)!}(\omega r)^{l+1}\;.
\end{eqnarray}

We now consider the up-modes solution to the radial equation with the boundary conditions given by Eq.\eqref{up_bc}. It is clear the in-mode solution and the up-mode solution are two indepedent solutions to the radial equation \eqref{radial_eq}. Then the Wronskian of the two solutions are independent of the radial coordinate. By substituting the asymptotic in-mode and up-mode solutions into the Wronskian
\begin{eqnarray}
    W(R^{\mathrm{in}}_{\omega l}(r^*),R^{\mathrm{up}}_{\omega l}(r^*))
    =R^{\mathrm{in}}_{\omega l}(r^*) \frac{d R^{\mathrm{up}}_{\omega l}(r)}{dr^*}-R^{\mathrm{up}}_{\omega l}(r^*) \frac{d R^{\mathrm{in}}_{\omega l}(r)}{dr^*}\;,
\end{eqnarray}
one can get the following relation by evaluating the Wronskian on the horizon and at the infinity 
\begin{eqnarray}
    A^{\mathrm{in}}_{\omega l}=B^{\mathrm{up}}_{\omega l}\;.
\end{eqnarray}
According to the discussion on the in-mode solution, we have 
\begin{eqnarray}
    A^{\mathrm{in}}_{\omega l}\approx A_-= \frac{(-i)^{l+1}2^{2l+2}(l!)^3(M\omega)^{l+1}}{(2l)!(2l+1)!}\;.
\end{eqnarray}
The above two equations gives us $B^{\mathrm{up}}_{\omega l}\sim (M\omega)^{l+1}\ll 1$, which means that at the low frequency, the wave propagating from the white hole is almost reflected back into the black hole by the potential barrier outside of the horizon. 

For up-mode solution in the far-field region, in order to cooperate the boundary condition, we should take the solution as 
\begin{eqnarray}\label{FF_up_sol}
    R_{FF}=B\sqrt{r} H^{(1)}_{l+\frac{1}{2}}(\omega r)\;,
\end{eqnarray}
where $H^{(1)}_{l+\frac{1}{2}}(\omega r)$ is Hankel function of the first kind. From the asymptotic property of the Hankel function, when $r\rightarrow \infty$, we have 
\begin{eqnarray}
    R_{FF}\rightarrow (-i)^{l+1} B \sqrt{\frac{2}{\pi\omega}} e^{i\omega r}\;. 
\end{eqnarray}
Compared with the boundary conditions of the up-modes, we can get the relation  
\begin{eqnarray}
    B=\sqrt{\frac{\pi\omega}{2}} \frac{2^{2l+2}(l!)^3(M\omega)^{l+1}}{(2l)!(2l+1)!}\;.
\end{eqnarray}
Substituting it back into the far-field solution \eqref{FF_up_sol} and using the small $r$ expansion of the Hankel function, we can get the approximated up-mode solution in the intermediate region as 
\begin{eqnarray}
   R^{\mathrm{up}}_{\omega l}(r)\approx -i \frac{2^{l+2}(l!)^2}{(2l+1)!}(M\omega) \left(\frac{M}{r}\right)^l\;.
\end{eqnarray}

\subsection{Decoherence rates in different vacua}

With the analytical expressions for the in-modes and the up-modes in hand, we can now calculate the decoherence rates for three kinds of vacua. The decoherence rate $\Gamma$ is related to kernel of the Wightman function $\mathcal{F}(\omega)$ at $\omega=0$. Here, $\omega=0$ should be understood as the limit $\omega\rightarrow 0$. Therefore, the dominant contribution to the Wightman function comes from the $l=0$ modes for the in-mode solution. While for the up-mode solution, due to $r_0\gg M$, the dominant contribution to the Wightman function also comes from the $l=0$ modes.  

For the Boulware vacuum, the low-frequency behavior of the Wightman two-point function is given by
\begin{eqnarray}
G_{\omega 0}^B(r,r')=4\omega^2\Theta(\omega) \left(4M^2+rr' \right)\;,
\end{eqnarray}
which scales quadratically with frequency in the infrared limit. Since  the decoherence rate at the long-time limit is controlled by the $\omega\to 0$ behavior of the field correlation function, substituting this expression into $\mathcal{F}(\omega)$ immediately gives 
\begin{eqnarray}
\Gamma_B=0\;.
\end{eqnarray}
Physically, this reflects the fact that the Boulware vacuum contains no thermal flux associated with the horizon, and therefore does not provide an efficient environmental channel to distinguish the two branches of the spatial superposition. As a result, quantum coherence is preserved in the long-time limit.

For the Unruh vacuum, the dominant low-frequency contribution takes the form
\begin{eqnarray}\label{U_G}
G_{\omega 0}^U(r,r')=
\frac{16M^2\omega^2}{1-e^{-8\pi M\omega}}
+4\omega^2\Theta(\omega) rr' \;.
\end{eqnarray}
Substituting this expression into $\mathcal{F}(\omega)$ and taking the $\omega\rightarrow 0$ limit yields
\begin{eqnarray}
\Gamma_U=\frac{\lambda^2 d^2 M}{4\pi^2 r_0^4}\;.
\end{eqnarray}
The nonvanishing decoherence rate originates from the thermal nature of the up-modes in the Unruh vacuum, which encode Hawking radiation and provide an environmental channel that can carry which-way information. It should be noted that the second term in Eq.~\eqref{U_G}, corresponding to the in-modes, scales as $\omega^2$ and is suppressed in the infrared limit. Consequently, the in-modes do not contribute to decoherence in the long-time regime.

For the Hartle-Hawking vacuum, the dominant contribution is given by 
\begin{eqnarray}
G_{\omega 0}^H(r,r')
=
\frac{4\omega^2}{1-e^{-8\pi M\omega}}
\left(4M^2 + rr'\right)\;.
\end{eqnarray}
This leads to the decoherence rate
\begin{eqnarray}
\Gamma_H=
\frac{\lambda^2 d^2 M}{4\pi^2 r_0^4}\;,
\end{eqnarray}
which coincides with the result in the Unruh vacuum. This agreement indicates that the dominant contribution to decoherence arises from the up-modes, which are common to both vacua. Although the Hartle-Hawking vacuum additionally contains thermal incoming radiation from infinity, the contribution from the in-modes remains infrared suppressed and therefore does not affect the long-time decoherence rate.

\section{Local calculation of decoherence in Reissner-Nordstr\"{o}m black holes}
\label{sec:rn}

In this section, we extend the local analysis of decoherence to the Reissner-Nordstr\"{o}m black hole spacetime.

For an Reissner-Nordstr"{o}m black hole, the metric function takes the form of 
\begin{eqnarray}
f(r)=1-\frac{2M}{r}+\frac{Q^2}{r^2}\;,
\end{eqnarray}
where $M$ and $Q$ denote the mass and charge of the black hole, respectively. Consider a massless scalar field in this background. After separating variables for the massless scalar field according to Eq.~\eqref{variable_separating}, one obtains the same radial equation as in the Schwarzschild case, except that the effective potential is modified to
\begin{eqnarray}
V(r)=\left(1-\frac{2M}{r}+\frac{Q^2}{r^2}\right)
\left(
\frac{l(l+1)}{r^2}
+\frac{2M}{r^3}
-\frac{2Q^2}{r^4}
\right)\;.
\end{eqnarray}

To determine the decoherence rate, we require analytical approximations to the mode functions. Following the same procedure as in the lase section, one can obtain the approximated mode solutions far from the horizon. For an Reissner-Nordstr"{o}m black hole, the asymptotic solutions were obtained in Ref.~\cite{Castineiras:2000qt} using the asymptotic matching method. In the low-frequency regime, the in- and up-mode solutions are approximately given by (see Eqs.~(2.28) and (2.33) in Ref.~\cite{Castineiras:2000qt})
\begin{eqnarray}
R_{\omega l}^{\mathrm{in}}(r)
&\approx&
\frac{i^{3l+3}2^{l+1}l!}{(2l+1)!}
(\omega r)^{l+1}\;,\\
R^{\mathrm{up}}_{\omega l}(r)
&\approx&
-i
\frac{2^{l+2}(l!)^2}{(2l+1)!}
\frac{\left(M+\sqrt{M^2-Q^2}\right)}{2M}
\left(\frac{\sqrt{M^2-Q^2}}{M}\right)^l
(M\omega)
\left(\frac{M}{r}\right)^l\;.
\end{eqnarray}
The in-mode solution remains unchanged from the Schwarzschild case, while the up-mode solution smoothly reduces to the Schwarzschild result in the limit $Q\to 0$, as expected.

We now evaluate the decoherence rates in the Boulware, Unruh, and Hartle-Hawking vacua. As in the Schwarzschild case, the long-time decoherence rate is controlled by the infrared behavior of the Wightman function, whose dominant contribution also arises from the $l=0$ sector of the up-modes. Consequently, one finds
\begin{eqnarray}
\Gamma_B=0\;,\qquad
\Gamma_U=
\Gamma_H=
\frac{\lambda^2 d^2
\left(M+\sqrt{M^2-Q^2}\right)^2}
{16\pi^2 M r_0^4}\;.
\end{eqnarray}

Several features are worth emphasizing. First, the decoherence rate vanishes in the Boulware vacuum, reflecting the absence of thermal horizon excitations. Second, the Unruh and Hartle-Hawking vacua yield the same decoherence rate, indicating that the dominant contribution originates from the thermal horizon modes common to both vacua. Similar to the Schwarzschild case, the contribution from the in-modes is suppressed in the infrared limit and therefore does not affect the long-time decoherence rate. This implies that decoherence is predominantly induced by the horizon-associated up-modes rather than by incoming radiation from spatial infinity.

At last, we should note that the decoherence rate in the Unruh and Hartle-Hawking vacua does not approach zero in the extremal limit $M=Q$. In our previous work \cite{Li:2024lfv}, we investigated the decoherence of a spatial superposition outside a Reissner-Nordstr\"{o}m black hole induced by the emission of low energy photons from a charged particle. There, it was shown that, due to the Meissner effect, electromagnetic fields are expelled from the event horizon in the extremal limit, leading to a strong suppression of decoherence. By contrast, in the present case where the particle is coupled to a scalar field, such a suppression is absent, and the decoherence rate remains finite even for the Reissner-Nordstr\"{o}m extremal black holes. This indicates that the decoherence behavior is also sensitive to the nature of the environmental field to which the source particle is coupled.

\section{Conclusion and discussion}
\label{sec:con_dis}

In this work, we investigated the decoherence of a spatial quantum superposition of a static particle induced by the scalar fields in Schwarzschild spacetime. Treating the particle as a localized source coupled to a scalar field, we reformulated decoherence through the generation of field coherent states and derived a local expression for the decoherence functional in terms of the Wightman two-point function. In the long-time limit, the decoherence dynamics is characterized by a local decoherence rate determined by the infrared behavior of field correlations.

Applying this framework to Schwarzschild and Reissner-Nordstr\"{o}m black holes, we analyzed the Boulware, Unruh, and Hartle-Hawking vacua. We found that the decoherence behavior depends sensitively on the quantum state of the black hole environment. In particular, no decoherence occurs in the Boulware vacuum, whereas nonvanishing decoherence arises in the Unruh and Hartle-Hawking vacua due to thermal fluctuations generated by the black hole horizon. These results highlight the important role of the quantum state of spacetime in determining the stability of quantum coherence near black holes.

The local formulation developed here provides a field theoretic perspective on environment induced decoherence near horizons. Several extensions deserve further study, including electromagnetic and gravitational environments, rotating and charged black holes in different asymptotic behavior.


\acknowledgments

The authors would like to thank Hongbao Zhang for useful discussions. R.L. would like to thank National Natural Science Foundation of China (No.12575059) and Shandong Provincial Natural Science Foundation (No.ZR2025MS10) for funding support. Z.X.M. would like to acknowledge support from the Shandong Provincial Natural Science Foundation under projects ZR2023LLZ015 and ZR2024LLZ012.


 \bibliographystyle{JHEP}
 \bibliography{biblio.bib}

@book{weinberg2015lectures,
  title={Lectures on Quantum Mechanics},
  author={Weinberg, Steven},
  year={2015},
  edition={2},
  publisher={Cambridge University Press}
}

@article{Hornberger:2008xkz,
    author = "Hornberger, Klaus",
    title = "{Introduction to decoherence theory}",
    eprint = "quant-ph/0612118",
    archivePrefix = "arXiv",
    doi = "10.1007/978-3-540-88169-8_5",
    journal = "Lect. Notes Phys.",
    volume = "768",
    pages = "221--276",
    year = "2009"
}

@article{Zurek:1991vd,
    author = "Zurek, Wojciech H.",
    title = "{Decoherence and the transition from quantum to classical}",
    eprint = "quant-ph/0306072",
    archivePrefix = "arXiv",
    doi = "10.1063/1.881293",
    journal = "Phys. Today",
    volume = "44N10",
    pages = "36--44",
    year = "1991"
}

@article{Zurek:2003zz,
    author = "Zurek, Wojciech Hubert",
    title = "{Decoherence, einselection, and the quantum origins of the classical}",
    eprint = "quant-ph/0105127",
    archivePrefix = "arXiv",
    doi = "10.1103/RevModPhys.75.715",
    journal = "Rev. Mod. Phys.",
    volume = "75",
    pages = "715--775",
    year = "2003"
}

@article{Schlosshauer:2003zy,
    author = "Schlosshauer, Maximilian",
    title = "{Decoherence, the Measurement Problem, and Interpretations of Quantum Mechanics}",
    eprint = "quant-ph/0312059",
    archivePrefix = "arXiv",
    doi = "10.1103/RevModPhys.76.1267",
    journal = "Rev. Mod. Phys.",
    volume = "76",
    pages = "1267--1305",
    year = "2004"
}

@book{Breuer2002,
  author    = {Breuer, Heinz-Peter and Petruccione, Francesco},
  title     = {The Theory of Open Quantum Systems},
  year      = {2002},
  publisher = {Oxford University Press},
  location  = {New York},
  isbn      = {9780198520634}
}

@book{Giulini1996,
  author    = {Domenico Giulini and Erich Joos and Claus Kiefer and
               Joachim Kupsch and Ion-Olimpiu Stamatescu and
               Hans Dieter Zeh},
  title     = {Decoherence and the Appearance of a Classical World in Quantum Theory},
  publisher = {Springer-Verlag},
  address   = {Berlin},
  year      = {1996}
}

@article{Bassi:2017szd,
    author = "Bassi, Angelo and Gro{\ss}ardt, Andr{\'e} and Ulbricht, Hendrik",
    title = "{Gravitational Decoherence}",
    eprint = "1706.05677",
    archivePrefix = "arXiv",
    primaryClass = "quant-ph",
    doi = "10.1088/1361-6382/aa864f",
    journal = "Class. Quant. Grav.",
    volume = "34",
    number = "19",
    pages = "193002",
    year = "2017"
}

@article{Anastopoulos:2021jdz,
    author = "Anastopoulos, Charis and Hu, Bei-Lok",
    title = "{Gravitational decoherence: A thematic overview}",
    eprint = "2111.02462",
    archivePrefix = "arXiv",
    primaryClass = "gr-qc",
    doi = "10.1116/5.0077536",
    journal = "AVS Quantum Sci.",
    volume = "4",
    number = "1",
    pages = "015602",
    year = "2022"
}

@article{Blencowe:2012mp,
    author = "Blencowe, M. P.",
    title = "{Effective Field Theory Approach to Gravitationally Induced Decoherence}",
    eprint = "1211.4751",
    archivePrefix = "arXiv",
    primaryClass = "quant-ph",
    doi = "10.1103/PhysRevLett.111.021302",
    journal = "Phys. Rev. Lett.",
    volume = "111",
    number = "2",
    pages = "021302",
    year = "2013"
}

@article{Pikovski:2013qwa,
    author = "Pikovski, Igor and Zych, Magdalena and Costa, Fabio and Brukner, Caslav",
    title = "{Universal decoherence due to gravitational time dilation}",
    eprint = "1311.1095",
    archivePrefix = "arXiv",
    primaryClass = "quant-ph",
    doi = "10.1038/nphys3366",
    journal = "Nature Phys.",
    volume = "11",
    pages = "668--672",
    year = "2015"
}

@article{Nesterov:2020exl,
    author = "Nesterov, Alexander I. and Berman, Gennady P. and Fern{\'a}ndez, Manuel A. Rodr{\'\i}guez and Wang, Xidi",
    title = "{Decoherence as a detector of the Unruh effect}",
    eprint = "2003.05014",
    archivePrefix = "arXiv",
    primaryClass = "gr-qc",
    reportNumber = "LA-UR-20-22227",
    doi = "10.1103/PhysRevResearch.2.043230",
    journal = "Phys. Rev. Res.",
    volume = "2",
    number = "4",
    pages = "043230",
    year = "2020"
}

@article{Xu:2023tdt,
    author = "Xu, Hao",
    title = "{Decoherence and thermalization of Unruh-DeWitt detector in arbitrary dimensions}",
    eprint = "2301.12381",
    archivePrefix = "arXiv",
    primaryClass = "hep-th",
    doi = "10.1007/JHEP03(2023)179",
    journal = "JHEP",
    volume = "03",
    pages = "179",
    year = "2023"
}

@article{Li:2026pxg,
    author = "Li, Ran and Man, Zhong-Xiao and Wang, Jin",
    title = "{Probing Unruh Effect from Enhanced Decoherence}",
    eprint = "2603.26121",
    archivePrefix = "arXiv",
    primaryClass = "gr-qc",
    month = "3",
    year = "2026"
}

@book{Birrell:1982ix,
    author = "Birrell, N. D. and Davies, P. C. W.",
    title = "{Quantum Fields in Curved Space}",
    doi = "10.1017/CBO9780511622632",
    isbn = "978-0-511-62263-2, 978-0-521-27858-4",
    publisher = "Cambridge University Press",
    address = "Cambridge, UK",
    series = "Cambridge Monographs on Mathematical Physics",
    year = "1982"
}

@article{Danielson:2022tdw,
    author = "Danielson, Daine L. and Satishchandran, Gautam and Wald, Robert M.",
    title = "{Black holes decohere quantum superpositions}",
    eprint = "2205.06279",
    archivePrefix = "arXiv",
    primaryClass = "hep-th",
    doi = "10.1142/S0218271822410036",
    journal = "Int. J. Mod. Phys. D",
    volume = "31",
    number = "14",
    pages = "2241003",
    year = "2022"
}

@article{Danielson:2022sga,
    author = "Danielson, Daine L. and Satishchandran, Gautam and Wald, Robert M.",
    title = "{Killing horizons decohere quantum superpositions}",
    eprint = "2301.00026",
    archivePrefix = "arXiv",
    primaryClass = "hep-th",
    doi = "10.1103/PhysRevD.108.025007",
    journal = "Phys. Rev. D",
    volume = "108",
    number = "2",
    pages = "025007",
    year = "2023"
}

@article{Danielson:2024yru,
    author = "Danielson, Daine L. and Satishchandran, Gautam and Wald, Robert M.",
    title = "{Local description of decoherence of quantum superpositions by black holes and other bodies}",
    eprint = "2407.02567",
    archivePrefix = "arXiv",
    primaryClass = "hep-th",
    doi = "10.1103/PhysRevD.111.025014",
    journal = "Phys. Rev. D",
    volume = "111",
    number = "2",
    pages = "025014",
    year = "2025"
}

@article{Gralla:2023oya,
    author = "Gralla, Samuel E. and Wei, Hongji",
    title = "{Decoherence from horizons: General formulation and rotating black holes}",
    eprint = "2311.11461",
    archivePrefix = "arXiv",
    primaryClass = "hep-th",
    doi = "10.1103/PhysRevD.109.065031",
    journal = "Phys. Rev. D",
    volume = "109",
    number = "6",
    pages = "065031",
    year = "2024"
}

@article{Wilson-Gerow:2024ljx,
    author = "Wilson-Gerow, Jordan and Dugad, Annika and Chen, Yanbei",
    title = "{Decoherence by warm horizons}",
    eprint = "2405.00804",
    archivePrefix = "arXiv",
    primaryClass = "hep-th",
    reportNumber = "CALT-TH 2024-017",
    doi = "10.1103/PhysRevD.110.045002",
    journal = "Phys. Rev. D",
    volume = "110",
    number = "4",
    pages = "045002",
    year = "2024"
}

@article{Biggs:2024dgp,
    author = "Biggs, Anna and Maldacena, Juan",
    title = "{Comparing the decoherence effects due to black holes versus ordinary matter}",
    eprint = "2405.02227",
    archivePrefix = "arXiv",
    primaryClass = "hep-th",
    month = "5",
    year = "2024"
}

@article{Li:2024guo,
    author = "Li, Ran",
    title = {{Decoherence of quantum superpositions by Reissner-Nordstr{\"o}m black holes}},
    eprint = "2411.04734",
    archivePrefix = "arXiv",
    primaryClass = "hep-th",
    doi = "10.1103/PhysRevD.111.024040",
    journal = "Phys. Rev. D",
    volume = "111",
    number = "2",
    pages = "024040",
    year = "2025"
}

@article{Li:2024lfv,
    author = "Li, Ran",
    title = "{Note on the local calculation of decoherence of quantum superpositions in de Sitter spacetime}",
    eprint = "2501.00213",
    archivePrefix = "arXiv",
    primaryClass = "hep-th",
    doi = "10.1103/PhysRevD.111.044022",
    journal = "Phys. Rev. D",
    volume = "111",
    number = "4",
    pages = "044022",
    year = "2025"
}

@article{Li:2025vcm,
    author = "Li, Ran and Man, Zhong-Xiao and Wang, Jin",
    title = {{Decoherence of quantum superpositions in near-extremal Reissner-Nordstr{\"o}m black holes with quantum gravity corrections}},
    eprint = "2505.07480",
    archivePrefix = "arXiv",
    primaryClass = "hep-th",
    doi = "10.1007/JHEP08(2025)079",
    journal = "JHEP",
    volume = "08",
    pages = "079",
    year = "2025"
}

@article{Kawamoto:2025kfu,
    author = "Kawamoto, Shoichi and Lee, Da-Shin and Yeh, Chen-Pin",
    title = "{Decoherence by black holes via holography}",
    eprint = "2505.17450",
    archivePrefix = "arXiv",
    primaryClass = "hep-th",
    doi = "10.1007/JHEP01(2026)154",
    journal = "JHEP",
    volume = "01",
    pages = "154",
    year = "2026"
}

@article{Fahn:2025fxl,
    author = "Fahn, Max Joseph and Pesci, Alessandro",
    title = "{Effects of quantum geometry on the decoherence induced by black holes}",
    eprint = "2507.16911",
    archivePrefix = "arXiv",
    primaryClass = "gr-qc",
    doi = "10.1103/279x-zgl1",
    journal = "Phys. Rev. D",
    volume = "112",
    number = "12",
    pages = "L121502",
    year = "2025"
}

@article{Fahn:2025jxh,
    author = "Fahn, Max Joseph and Pesci, Alessandro",
    title = "{Horizon quantum geometries and decoherence}",
    eprint = "2507.18709",
    archivePrefix = "arXiv",
    primaryClass = "gr-qc",
    doi = "10.1103/qhd4-pj8w",
    journal = "Phys. Rev. D",
    volume = "112",
    number = "12",
    pages = "124036",
    year = "2025"
}

@article{Batista:2026otr,
    author = "Batista, Levy B. N. and Landulfo, Andr{\'e} G. S. and Mann, Robert B. and Matsas, George E. A.",
    title = "{Nonperturbative Danielson-Satishchandran-Wald Decoherence with Unruh-DeWitt detectors}",
    eprint = "2605.00956",
    archivePrefix = "arXiv",
    primaryClass = "gr-qc",
    month = "5",
    year = "2026"
}

@article{Biggs:2026zlp,
    author = "Biggs, Anna and Trezzi, Stefano",
    title = "{Not all black holes decohere quantum superpositions}",
    eprint = "2605.23880",
    archivePrefix = "arXiv",
    primaryClass = "hep-th",
    month = "5",
    year = "2026"
}

@article{Danielson:2025iud,
    author = "Danielson, Daine L. and Kudler-Flam, Jonah and Satishchandran, Gautam and Wald, Robert M.",
    title = "{How to minimize the decoherence caused by black holes}",
    eprint = "2501.04773",
    archivePrefix = "arXiv",
    primaryClass = "hep-th",
    doi = "10.1103/67vv-km43",
    journal = "Phys. Rev. D",
    volume = "112",
    number = "2",
    pages = "025012",
    year = "2025"
}

@article{Kudler-Flam:2025yur,
    author = "Kudler-Flam, Jonah and Penington, Geoff",
    title = "{It costs nothing to teleport information into a black hole}",
    eprint = "2504.01058",
    archivePrefix = "arXiv",
    primaryClass = "hep-th",
    doi = "10.1142/S0218271825430023",
    journal = "Int. J. Mod. Phys. D",
    volume = "34",
    number = "16",
    pages = "2543002",
    year = "2025"
}

@article{Danielson:2025aji,
    author = "Danielson, Daine L. and Satishchandran, Gautam",
    title = "{Horizons and Soft Quantum Information}",
    eprint = "2512.20754",
    archivePrefix = "arXiv",
    primaryClass = "hep-th",
    reportNumber = "MIT-CTP/5990",
    month = "12",
    year = "2025"
}

@article{Fahn:2026rei,
    author = "Fahn, Max Joseph and Pesci, Alessandro",
    title = "{On the nature of entangling photons in horizon-induced decoherence}",
    eprint = "2605.19588",
    archivePrefix = "arXiv",
    primaryClass = "gr-qc",
    month = "5",
    year = "2026"
}

@article{Gallock-Yoshimura:2026vku,
    author = "Gallock-Yoshimura, Kensuke and Hasegawa, Yoshihiko",
    title = "{Trade-off Relation for Black Hole Entropy Fluctuations}",
    eprint = "2605.26214",
    archivePrefix = "arXiv",
    primaryClass = "gr-qc",
    month = "5",
    year = "2026"
}

@book{Wald:1995yp,
    author = "Wald, Robert M.",
    title = "{Quantum Field Theory in Curved Space-Time and Black Hole Thermodynamics}",
    isbn = "978-0-226-87027-4",
    publisher = "University of Chicago Press",
    address = "Chicago, IL",
    series = "Chicago Lectures in Physics",
    year = "1995"
}

@article{Boulware:1974dm,
    author = "Boulware, David G.",
    title = "{Quantum Field Theory in Schwarzschild and Rindler Spaces}",
    reportNumber = "RLO-1388-683",
    doi = "10.1103/PhysRevD.11.1404",
    journal = "Phys. Rev. D",
    volume = "11",
    pages = "1404",
    year = "1975"
}

@article{Unruh:1976db,
    author = "Unruh, W. G.",
    title = "{Notes on black hole evaporation}",
    doi = "10.1103/PhysRevD.14.870",
    journal = "Phys. Rev. D",
    volume = "14",
    pages = "870",
    year = "1976"
}

@article{Hartle:1976tp,
    author = "Hartle, J. B. and Hawking, S. W.",
    title = "{Path Integral Derivation of Black Hole Radiance}",
    doi = "10.1103/PhysRevD.13.2188",
    journal = "Phys. Rev. D",
    volume = "13",
    pages = "2188--2203",
    year = "1976"
}

@book{Scully:1997hcl,
    author = "Scully, Marlan O. and Zubairy, M. Suhail",
    title = "{Quantum Optics}",
    doi = "10.1017/cbo9780511813993",
    publisher = "Cambridge University Press",
    month = "9",
    year = "1997"
}

@article{Glauber:1963tx,
    author = "Glauber, Roy J.",
    title = "{Coherent and incoherent states of the radiation field}",
    doi = "10.1103/PhysRev.131.2766",
    journal = "Phys. Rev.",
    volume = "131",
    pages = "2766--2788",
    year = "1963"
}

@ARTICLE{2012JPhA...45x4002S,
       author = {{Sanders}, Barry C.},
        title = "{Review of entangled coherent states}",
      journal = {Journal of Physics A Mathematical General},
         year = 2012,
        month = jun,
       volume = {45},
       number = {24},
          eid = {244002},
        pages = {244002},
          doi = {10.1088/1751-8113/45/24/244002},
archivePrefix = {arXiv},
       eprint = {1112.1778},
 primaryClass = {quant-ph},
       adsurl = {https://ui.adsabs.harvard.edu/abs/2012JPhA...45x4002S}
}

@article{Zhang:1990fy,
    author = "Zhang, Wei-Min and Feng, Da Hsuan and Gilmore, Robert",
    title = "{Coherent States: Theory and Some Applications}",
    doi = "10.1103/RevModPhys.62.867",
    journal = "Rev. Mod. Phys.",
    volume = "62",
    pages = "867--927",
    year = "1990"
}

@article{Das:1996we,
    author = "Das, Sumit R. and Gibbons, Gary W. and Mathur, Samir D.",
    title = "{Universality of low-energy absorption cross-sections for black holes}",
    eprint = "hep-th/9609052",
    archivePrefix = "arXiv",
    reportNumber = "TIFR-TH-96-49, MIT-CTP-2565",
    doi = "10.1103/PhysRevLett.78.417",
    journal = "Phys. Rev. Lett.",
    volume = "78",
    pages = "417--419",
    year = "1997"
}

@article{Gubser:1996xe,
    author = "Gubser, S. S. and Klebanov, Igor R.",
    title = "{Emission of charged particles from four-dimensional and five-dimensional black holes}",
    eprint = "hep-th/9608108",
    archivePrefix = "arXiv",
    reportNumber = "PUPT-1644",
    doi = "10.1016/S0550-3213(96)00496-8",
    journal = "Nucl. Phys. B",
    volume = "482",
    pages = "173--186",
    year = "1996"
}

@article{Kanti:2002nr,
    author = "Kanti, Panagiota and March-Russell, John",
    title = "{Calculable corrections to brane black hole decay. 1. The scalar case}",
    eprint = "hep-ph/0203223",
    archivePrefix = "arXiv",
    reportNumber = "CERN-TH-2002-014",
    doi = "10.1103/PhysRevD.66.024023",
    journal = "Phys. Rev. D",
    volume = "66",
    pages = "024023",
    year = "2002"
}

@article{Belenchia:2018szb,
    author = "Belenchia, Alessio and Wald, Robert M. and Giacomini, Flaminia and Castro-Ruiz, Esteban and Brukner, \v{C}aslav and Aspelmeyer, Markus",
    title = "{Quantum Superposition of Massive Objects and the Quantization of Gravity}",
    eprint = "1807.07015",
    archivePrefix = "arXiv",
    primaryClass = "quant-ph",
    doi = "10.1103/PhysRevD.98.126009",
    journal = "Phys. Rev. D",
    volume = "98",
    number = "12",
    pages = "126009",
    year = "2018"
}

@article{Danielson:2021egj,
    author = "Danielson, Daine L. and Satishchandran, Gautam and Wald, Robert M.",
    title = "{Gravitationally mediated entanglement: Newtonian field versus gravitons}",
    eprint = "2112.10798",
    archivePrefix = "arXiv",
    primaryClass = "quant-ph",
    doi = "10.1103/PhysRevD.105.086001",
    journal = "Phys. Rev. D",
    volume = "105",
    number = "8",
    pages = "086001",
    year = "2022"
}

@article{Scully:2017utk,
    author = "Scully, Marlan O. and Fulling, Stephen and Lee, David and Page, Don N. and Schleich, Wolfgang and Svidzinsky, Anatoly",
    title = "{Quantum optics approach to radiation from atoms falling into a black hole}",
    eprint = "1709.00481",
    archivePrefix = "arXiv",
    primaryClass = "quant-ph",
    doi = "10.1073/pnas.1807703115",
    journal = "Proc. Nat. Acad. Sci.",
    volume = "115",
    number = "32",
    pages = "8131--8136",
    year = "2018"
}

@article{Shallue:2025zto,
    author = "Shallue, Christopher J. and Carroll, Sean M.",
    title = "{What Hawking radiation looks like as you fall into a black hole}",
    eprint = "2501.06609",
    archivePrefix = "arXiv",
    primaryClass = "gr-qc",
    doi = "10.1103/y7kj-4zjw",
    journal = "Phys. Rev. D",
    volume = "112",
    number = "8",
    pages = "085013",
    year = "2025"
}

@article{Fabbri:1975sa,
    author = "Fabbri, R.",
    title = "{Scattering and absorption of electromagnetic waves by a Schwarzschild black hole}",
    doi = "10.1103/PhysRevD.12.933",
    journal = "Phys. Rev. D",
    volume = "12",
    pages = "933--942",
    year = "1975"
}

@article{Candelas:1980zt,
    author = "Candelas, P.",
    title = "{Vacuum Polarization in Schwarzschild Space-Time}",
    doi = "10.1103/PhysRevD.21.2185",
    journal = "Phys. Rev. D",
    volume = "21",
    pages = "2185--2202",
    year = "1980"
}

@article{Christensen:1977jc,
    author = "Christensen, S. M. and Fulling, S. A.",
    title = "{Trace Anomalies and the Hawking Effect}",
    doi = "10.1103/PhysRevD.15.2088",
    journal = "Phys. Rev. D",
    volume = "15",
    pages = "2088--2104",
    year = "1977"
}

@article{Hodgkinson:2014iua,
    author = "Hodgkinson, Lee and Louko, Jorma and Ottewill, Adrian C.",
    title = "{Static detectors and circular-geodesic detectors on the Schwarzschild black hole}",
    eprint = "1401.2667",
    archivePrefix = "arXiv",
    primaryClass = "gr-qc",
    doi = "10.1103/PhysRevD.89.104002",
    journal = "Phys. Rev. D",
    volume = "89",
    number = "10",
    pages = "104002",
    year = "2014"
}

@article{Ng:2014kha,
    author = "Ng, Keith K. and Hodgkinson, Lee and Louko, Jorma and Mann, Robert B. and Martin-Martinez, Eduardo",
    title = "{Unruh-DeWitt detector response along static and circular geodesic trajectories for Schwarzschild-AdS black holes}",
    eprint = "1406.2688",
    archivePrefix = "arXiv",
    primaryClass = "quant-ph",
    doi = "10.1103/PhysRevD.90.064003",
    journal = "Phys. Rev. D",
    volume = "90",
    number = "6",
    pages = "064003",
    year = "2014"
}

@article{Caribe:2023fhr,
    author = "Carib{\'e}, Jo{\~a}o G. A. and Jonsson, Robert H. and Casals, Marc and Kempf, Achim and Mart{\'\i}n-Mart{\'\i}nez, Eduardo",
    title = "{Lensing of vacuum entanglement near Schwarzschild black holes}",
    eprint = "2303.01402",
    archivePrefix = "arXiv",
    primaryClass = "quant-ph",
    reportNumber = "NORDITA 2023-022",
    doi = "10.1103/PhysRevD.108.025016",
    journal = "Phys. Rev. D",
    volume = "108",
    number = "2",
    pages = "025016",
    year = "2023"
}

@article{Castineiras:2000qt,
    author = "Castineiras, J. and Matsas, George E. A.",
    title = "{Low-energy sector quantization of a massless scalar field outside a Reissner-Nordstrom black hole and static sources}",
    eprint = "gr-qc/0002072",
    archivePrefix = "arXiv",
    reportNumber = "IFT-P-020-2000",
    doi = "10.1103/PhysRevD.62.064001",
    journal = "Phys. Rev. D",
    volume = "62",
    pages = "064001",
    year = "2000"
}

\end{document}